# Encryption of Audio Signals Using the Elzaki Transformation and the Lorenz Chaotic System

Shadman R. Kareem

Department of Mathematics, Faculty of Science and Health, Koya University, Erbil. Iraq
Department of Computer Science, College of Information Technology and Computer Sciences, Catholic University in Erbil. Iraq

**Abstract**

The preservation of image privacy during storage and transmission is of paramount importance in several areas including healthcare, military, safe communication, and video conferencing. Protecting data privacy demands the use of robust image encryption techniques. Several cryptographic techniques have been particularly designed to ensure the privacy of digital images. This study presents a novel method for encrypting color images utilizing chaos theory and a special transformation. This indicated approach first employs the Lorenz chaos theory to scramble the audio files. Following that, we utilize a technique that involves using the Maclaurin series expansion of hyperbolic functions and the Elzaki transform to encrypt the audio. Subsequently, we decode it by applying the inverse Elzaki transform. The key for the coefficients obtained from the transformation is created using modular arithmetic methods. Comparisons between the techniques are conducted based on a number of performance measures, including entropy analysis, spectrogram plotting, and correlation coefficients. Theoretical analysis and simulation indicate the efficacy of the proposed approach and confirm that this method is suitable for actual audio encryption. Moreover, the security inquiry indicates that an extra layer of security is provided by the provided audio encryption approach.

## 1. Introduction

Individuals worldwide are currently required to transmit large amounts of digital data over networks that lack security. Transmitting private and personal images and audio over an unsecured network always exposes them to the potential for unauthorized access, illustrating the critical importance of addressing security considerations when exchanging information. People have traditionally regarded cryptography as one of the most efficient techniques for ensuring the safe transmission of information. It protects confidentiality and integrity by employing a variety of encryption methods.

Conventional encryption methods notably AES, DES, and RSA are deemed insufficient for A encryption due to their higher processing time requirements. In addition, the increase in multimedia data often results in substantial pixel correlation, rendering conventional encryption methods such as AES, DES, and RSA insufficient for attaining optimal data efficiency during transmission. In order to address these challenges, a variety of audio techniques for encryption have been indicated, with just a few of them based on transformation[1,2], applying discrete wavelet transform. In addition, in[3], this study investigates an innovative approach for encrypting audio that utilizes the combined capabilities of Laplace transforms and hyperbolic functions. The suggested approach offers an original technique for converting audio data into a mathematically obscured format, rendering it unintelligible to anybody without the decryption key. Digital content often use non-traditional encryption techniques to improve both speed and security. These unconventional methods include a variety of techniques for altering and restructuring data, with some ones relying on visual alterations.

The combined use of chaos theory with encryption in current technologies presents a new framework for ensuring data security. Researchers are very intrigued by these improvements as they strive to develop novel cryptography methods. The unique characteristics of chaos theory have led to its widespread use in the encryption of pictures and audio. Currently, several techniques for encryption that utilize chaos are in development. In[4] this work presents a new encryption method that use the chaotic standard map to generate a diffusion effect during its replacement phase. The suggested technique leverages the inherent properties of chaos, such as its sensitivity to beginning circumstances and pseudo-randomness, to provide secure encryption. In[5] this work presents an innovative method to encrypting images by using an enhanced Lorenz system to enhance both the security level and the efficiency of the encryption process. This study presents a novel approach to enhance the nonlinear kinetic complexity and ergodicity of the Lorenz system. Key streams are used to execute pixel permutation and diffusion operations, ensuring an elevated level of security. Key streams are utilized for carrying out pixel permutation and diffusion operations, offering an excellent level of security. We utilize chaos to amplify the dispersion and perplexity in pictures. Chaotic maps have several benefits, such as a vast array of potential keys and a significant level of security. considering its inherent stochasticity and indeterminacy, chaos theory is often used in the encryption of audio and visual data.

While chaos-based encryption creates extremely random numbers, it is inherently complicated and unpredictable, that has attracted a lot of curiosity in its application in recent years[6]. Studies encourage the use of chaotic systems as sources of entropy for producing true random numbers, in both discrete and continuous time. True random numbers generators that utilize chaos are obviously favored by researchers. A secure communication system and several kinds of cryptographic applications need a supply of random numbers. The accuracy and unpredictability that are generated random numbers influence the security and reliability of these systems[7].

When creating chaotic systems random number generators that need an entropy source, the Lorenz system is a common option. The Lorenz system is an autonomous three-dimensional system with an odd attractor that is sensitive to beginning circumstances. Under certain differential equation parameter combinations, it exhibits chaotic behavior[8]. Because of its intrinsic unpredictability, the Lorenz system is a great option for applications that need to generate random numbers. The Lorenz system, in its basic form[9], its application to a field-programmable gate array with discrete time[10], and an improved version of the system[11], has been used in many investigations. The Lorenz system is especially attractive for cryptographic applications, such picture encryption[12], initial value maximization[13], and safe random bit generation[14], due to its very flexible and intricate features.

By integrating multiple approaches, it is conceivable to develop more robust security modules that may effectively address security challenges and provide dependable outcomes in the encryption process. This method employs a variety of strategies to improve the security and reliability of encryption processing systems. Researchers may create a comprehensive framework that combines cryptography, chaotic systems, and advanced image processing techniques to improve security and prevent unauthorized access and tampering. The experimental results highlight the efficacy of this integrated method, demonstrating significant improvements in both security and computing efficiency. Utilizing frameworks like this is especially crucial for applications that need real-time data handling while maintaining an elevated level of data security, including encrypted connections and secret information transmission.

In[15], their approach combines chaotic theory with a convolutional neural network (CNN) to provide sophisticated cryptography algorithms that contribute to improved security and resilience in image encryption. The strong feature extraction and transformation abilities of a CNN, along with the randomness and predictability of chaotic theory, make encryption techniques a lot harder to break and more complicated. For the purpose of to boost security, he suggests an inventive method of encrypting pictures in[16] that combines the symmetric RSA algorithm with the Arnold transform. They further confuse matters by using the Arnold transform to confuse the pixel coordinates within the image. The scrambled image is subsequently encrypted utilizing the RSA technique, that provides outstanding safety through its asymmetric key structure. Their work presents a new approach to audio signal encryption that utilizes the chaotic Hénon map with lifting wavelet transforms. The recommended approach utilizes the lifting wavelet transform to break down the audio signal into several frequency components, thus enhancing the efficiency and resilience of the encryption structure. Then, we employ the chaotic character of the Hénon map to encode and transform data[17].

The Elzaki transform, a contemporary integral transform analogous to the Laplace and Fourier transforms, plays a crucial role in resolving differential and integral equations[18,19] . The Elzaki transform, named after mathematician Taha Elzaki, has received considerable recognition for its effectiveness in handling many mathematical and engineering problems, especially in the discipline of signal processing. The Elzaki Transform was recently used in cryptography. This study presents an innovative approach to image encryption using the Elzaki transformation and substitution procedure in[20]. The Maclaurin series coefficient expansion enables the implementation of this methodology. Then encrypt the image using an infinite series of hyperbolic functions and the Elzaki transform, and decrypt it using the inverse Elzaki transform. The Elzaki transform is used in audio encryption to boost the efficiency of techniques for encryption through transforming audio signals into a modified domain predicated on their mathematical characteristics. This approach improves the audio data's resilience towards unauthorized manipulation and cryptographic assaults, hence optimizing security.

This study explores the present implementation of the Elzaki Transform in audio encryption, with the objective of reaching a balance between providing security while preserving the high-quality audio for communication purposes. This study introduces a novel approach that integrates chaotic picture encryption with the Elzaki transform for improving both the level of security and the general efficiency.  The procedure first encrypts the audio data, typically is represented as a vector, and then utilizes the naturally chaotic and nonlinear mapping characteristics of the Lorenz system to produce sequences that are unexpected. Nevertheless, despite these diligent efforts, the statistical analysis and computation of correlation coefficients suggest that just emphasizing on this pre-encryption phase does not provide enough protection.

The second part of the algorithm tries to get around this problem by using the Elzaki transform, along with a replacement method made possible by the expansion of Maclaurin series coefficients. This method entails encrypting the picture with an unending sequence of hyperbolic functions. Implementing the Elzaki Transform in this way greatly enhances the intricacy and resilience of the encryption procedure, guaranteeing heightened security for the audio data.

## 2. Lorenz Chaotic System

The Lorenz system, first introduced by Edward Lorenz in 1963, is a classic example of a chaotic system characterized by sensitive dependence on initial conditions. This system has found applications beyond its original context in atmospheric convection, including cryptography and secure communication. In this study, we explore the use of the Lorenz chaotic system for encrypting audio signals. The following set of nonlinear ordinary differential equations describes the Lorenz system:

$$\begin{aligned} \dot{x} &= \sigma(y - x), \\ \dot{y} &= x(\rho - z) - y, \\ \dot{z} &= xy - \beta z. \end{aligned} \quad (2)$$

where $x, y$, and $z$ are the state variables, and $\sigma, \rho$, and $\beta$ are the parameters of the system with the initial condition $(x_0, y_0, z_0) = (0.02, 0.02, 0.02)$. The behavior of the Lorenz system exhibits a strange attractor in phase space, which is highly sensitive to initial conditions and shows chaotic behavior, see the Figures (1-5). For more detail see the reference [3].

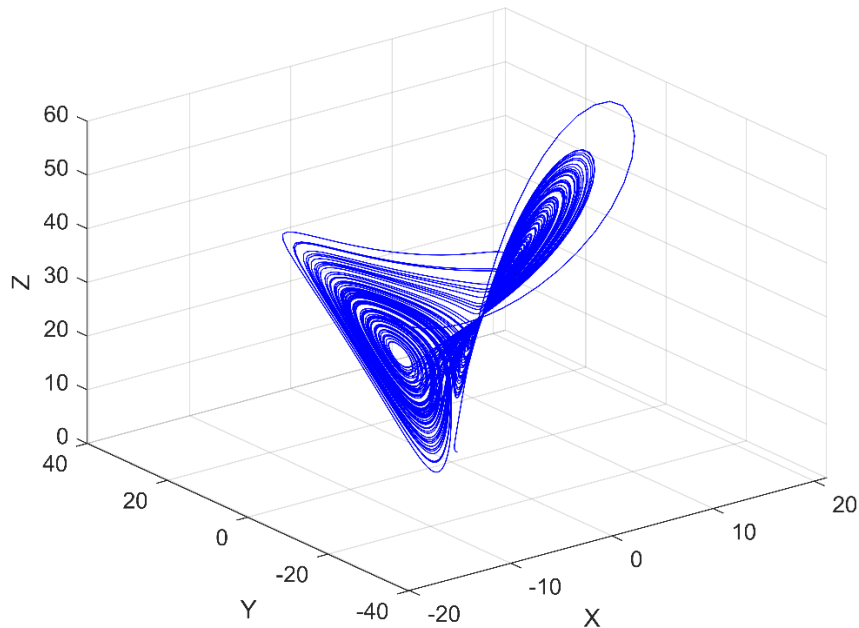

**Figure 1:** 3D phase portrait of the Lorenz chaotic system.

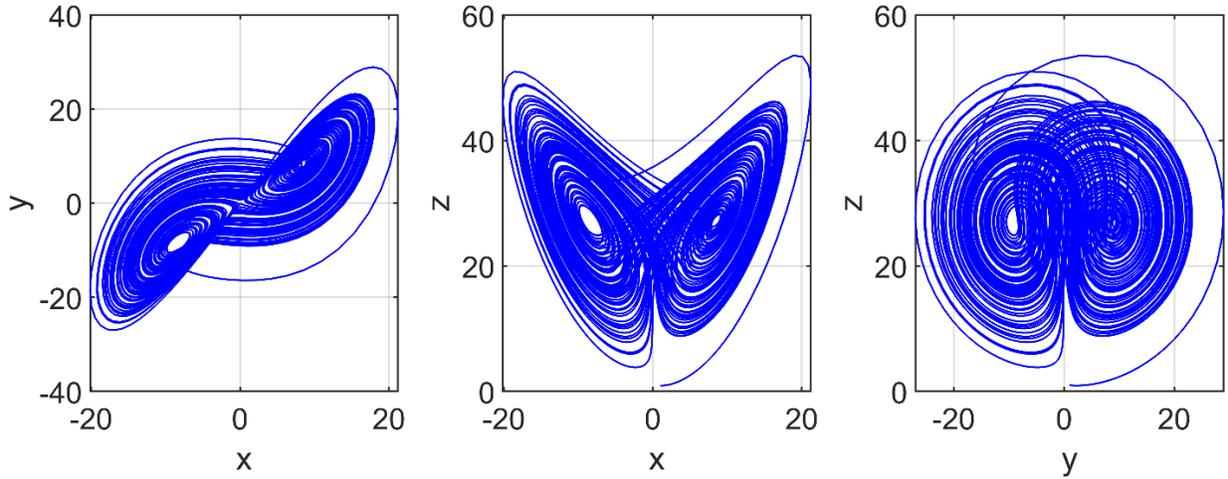

**Figure 2:** 2D phase portrait of the Lorenz chaotic system.

The chaotic behavior of the Lorenz system makes it suitable for cryptography, particularly audio encryption. The Lorenz system is good for audio encryption because it is sensitive to initial conditions. This means that even small changes in the starting parameters can have big effects on the outputs, as shown in Figures 3–5. This presents a challenge for unauthorized individuals attempting to duplicate the accurate decryption key without possessing specific understanding of the original situations. The output of the system is disorderly and seems to be without any identifiable patterns, which improves security. In addition, the system's extremely nonlinear structure makes it difficult to anticipate its output using simple linear transformations, which eventually boosts security.

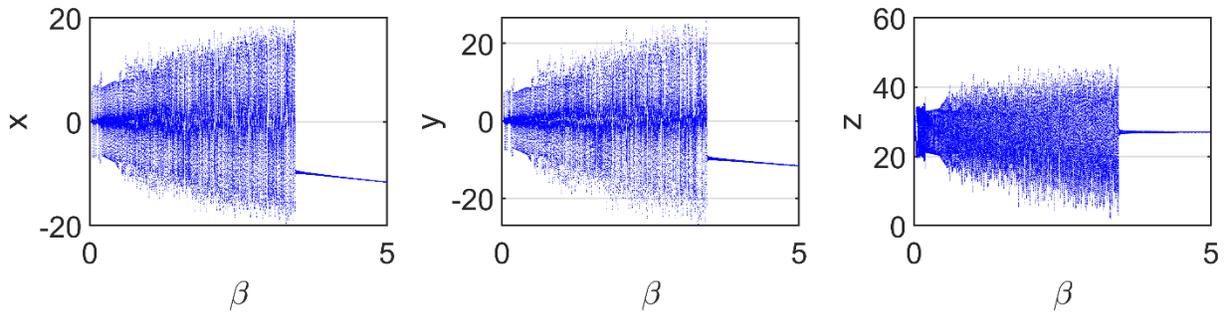

**Figure 3:** Bifurcation diagram for the Lorenz system, where $\sigma = 10$, $\rho = 28$, and $\beta \in [0, 5]$.

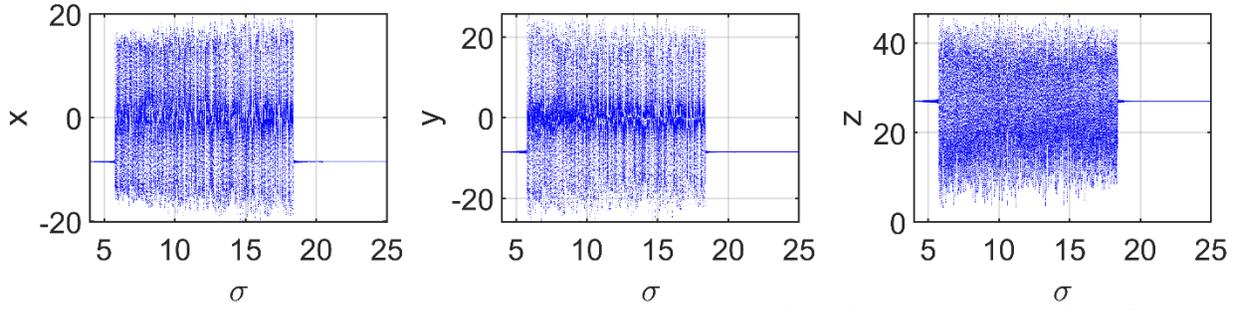
**Figure 4:** Bifurcation diagram for the Lorenz system, where $\beta = 8/3$, $\rho = 28$, and $\sigma \in [4, 25]$.

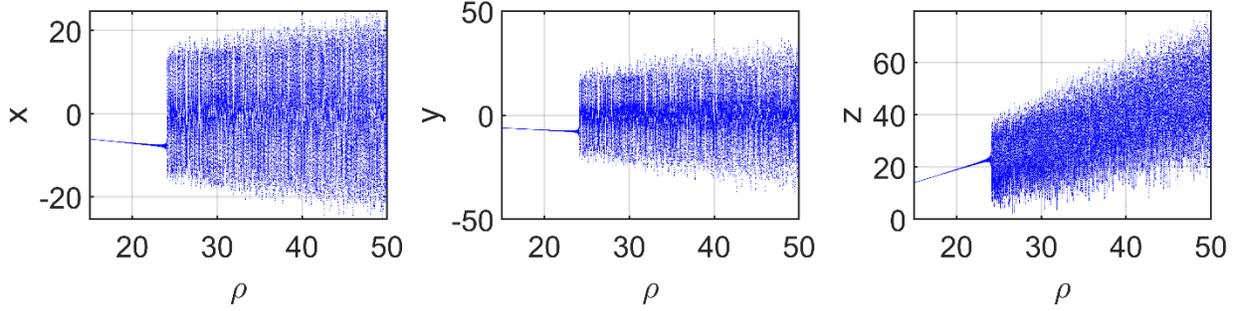
**Figure 5:** Bifurcation diagram for the Lorenz system, where $\sigma = 10$, $\beta = 8/3$, and $\rho \in [15, 50]$.

## 3. The Elzaki Transform

The Elzaki transformation, first presented by Tarig M. Elzaki in 2011, is an important transformation. The Elzaki transform has achieved considerable recognition in feasible mathematics and engineering thanks to its remarkable transforming properties, especially possess significant applications in signal processing and cryptography. This study explores the use of the Elzaki transform for encrypting audio files. The extensive application and impact of this approach are shown by its capacity to address an extensive variety of mathematical and technical issues. The application of the Elzaki Transform for encrypting audio signals demonstrates its flexibility in unconventional conditions. The Elzaki transform is a powerful tool for improving signal examination and modification in this specific situation. The audio encryption system is very desirable considering its unique features and capacity to adapt, providing robust security and fast processing[18].

The specified functions through set A are considered as follows:

$$A = \left\{ f(t) : \exists M, k_1, k_2 > 0, |f(t)| < M e^{\frac{|t|}{k_j}}, if\ t \in (-1)^j \times [0, \infty) \right\},$$

For a particular function in the set, M must have a finite value, but $k_1$ and $k_2$ might be either finite or infinite. The Elzaki Transformation may be mathematically represented in the following manner:

$$E[f(t)] = T(s) = s \int_0^\infty f(t) e^{\frac{-t}{s}} dt, k_1 \leq s \leq k_2, t \geq 0. \tag{1}$$

The following provides explanations of the Elzaki transformation and the inverse Elzaki transform for some basic functions:

- $E(k) = ks^2$ where $a$ is constant.
- $E(t^m) = m! \, s^{m+2}$,
- $E^{-1}(s^2) = 1$,
- $E^{-1}(s^{m+2}) = t^n/m!$.

## 4. Proposed Method

The following stages outline the procedural steps that enhance data security and privacy in the proposed hybrid mode of Elzaki transformation and the Lorenz Chaotic System, using them in cryptography and encryption technology.

### 4.1 Encryption Algorithm

To construct the Lorenz chaotic key, begin by setting up the Lorenz system (2) with specified parameters $\sigma = 10$, $\rho = 28$, and $\beta = 8/3$, as well as beginning with the initial condition $(x_0, y_0, z_0) = (0.02, 0.02, 0.02)$. Then, we employ Euler's technique for finding the numerical integration, to solve the Lorenz differential equations over a certain number of time steps $dt = 0.01$. The time series data for $x, y$, and $z$ will display chaotic behavior. These variables, usually denoted as $x$-component, $y$-component, and $z$-component of the Lorenz Chaotic Key. The diagram in Figure (6) depicts the successive processes of the proposed audio encryption method.

We consider the Maclaurin series of $t^2 e^{at}$ to be:

$$t^2 e^{at} = \sum_{n=0}^{\infty} \frac{a^n t^{n+2}}{n!} \tag{3}$$

For simulation trials, the proposed approach uses an "audio.wav" file as the carrier input. We executed the proposed methodology using MATLAB R2021a, employing two audio files (.wav) as input data. Figure 7 illustrate the behavior of the technique during the encryption and decryption operations. The Matlab function [**X**,Fs]=audioread('audio1.wav') retrieves the vector **X**, which represents our sample voice's amplitude values per second. This vector contains the message that requires encryption.

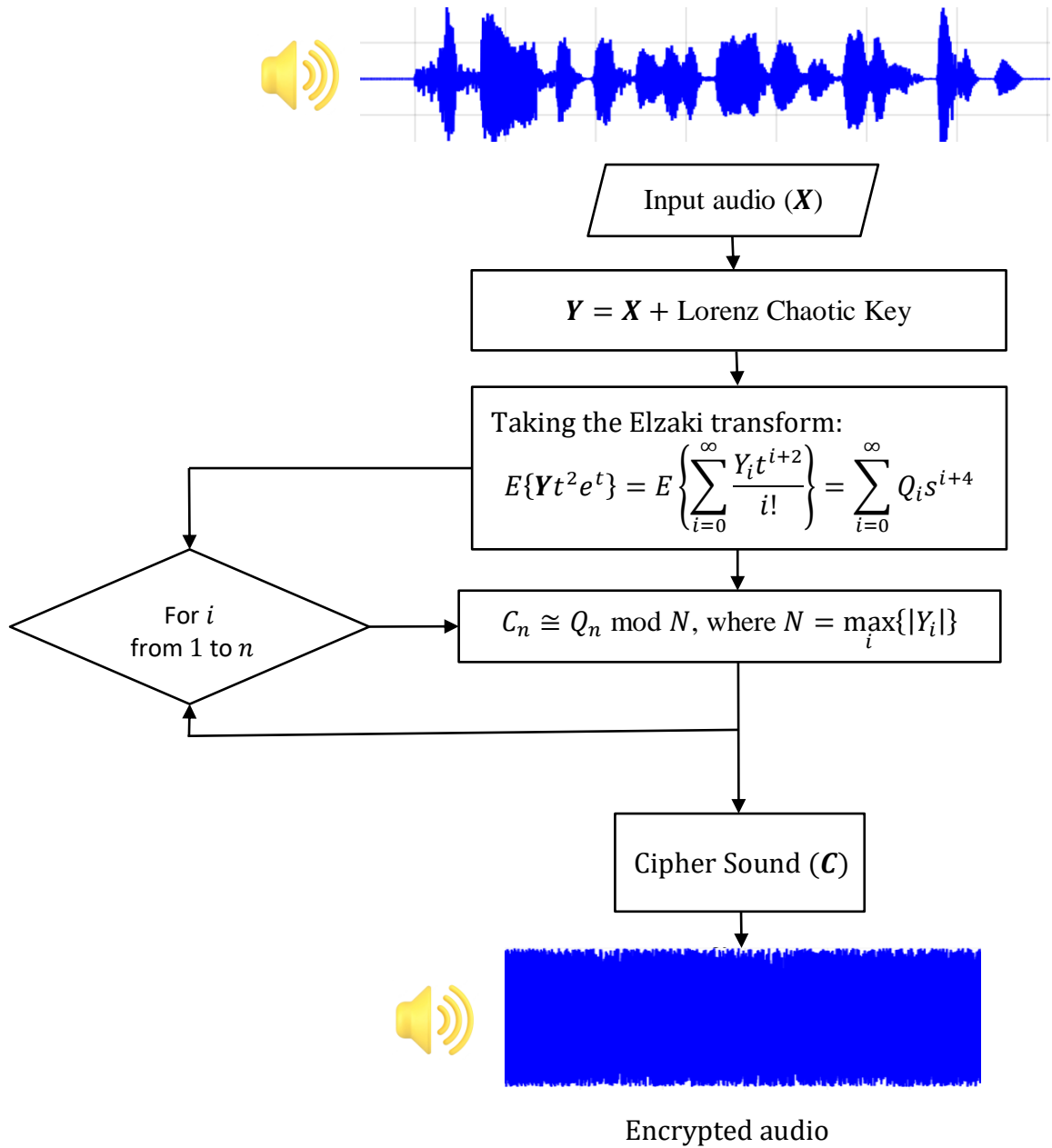

**Figure 6**: Diagram illustrating the sequence of steps in the audio encryption method being proposed.

**Step 1:** Inputting and reading audio data as a vector $X$.

**Step 2:** The pre-encoded audio $Y = X +$ Lorenz Chaotic Key.

**Step 3:** Make the entries $Y_n$ of the vector $Y$ for the audio data as the coefficients of the Maclaurin series of $t^2 e^t$ as follows:

$$Yt^2 e^t = \sum_{n=0}^{\infty} \frac{Y_n}{n!} t^{n+2}, \text{ where } X_n = 0, \text{ for } n > \text{length}(Y). \qquad (4)$$

**Step 4:** Utilizing the Elzaki transformation for Equation (4), we derive the transformed expression, as follows:

$$E\left[\sum_{n=0}^{\infty}\frac{Y_n}{n!}t^{n+2}\right] = s\int_0^\infty \sum_{n=0}^{\infty}\frac{Y_n}{n!}t^{n+2} e^{\frac{-t}{s}}dt = \sum_{n=0}^{\infty} Y_n(n^2+3n+2)s^{j+4} = \sum_{n=0}^{\infty} Q_n s^{j+4}, \quad (5)$$

where $Q_n = Y_n(n^2 + 3n + 2)$, for $n = 0,1,2, \ldots, \text{length}(Y)$.

**Step 5:** Find $C_n$ such that $C_n \cong Q_n \mod N$, where $N = \max_i\{|Y_i|\}$, for $n = 0,1,2, \ldots, \text{length}(Y)$. Then, the cipher audio become $C$.

Note that, the pre-encoded audio $Y$ in terms of $n$, for $n = 0,1,2, \ldots, \text{length}(Y)$, under the Elzaki transform of $Yt^2e^t$ can be converted to cipher audio $C$, where the components of this matrix are given by

$$C_n = Q_n - NK_n, \quad (6)$$

and

$$Q_n = (2n+1)Y_n, \quad (7)$$

with the modular-key

$$K_n = \frac{Q_n - C_n}{N}, \quad (8)$$

## 4.2 Decryption Algorithm

Decryption is the inverse procedure of encryption. During the decryption process, the input consists of the encrypted audio, which is represented as a vector denoted $C$. We use two confidential keys: the Lorenz Chaotic Key's $x$-component, $y$-component, and $z$-component, as well as the modular key (represented as $K$ in Equation (8)). We use these keys throughout the procedure, ultimately leading to the audio decryption. The decryption procedure includes a series of sequential processes with the objective of recovering the original audio from its encrypted state.

**Step 1:** Choose the encoded audio.

**Step 2:** Find $Q_n$ such that $Q_n \cong C_n \mod N$, for $n = 0,1,2, \ldots, \text{length}(X)$. Then, make it as the coefficient of the follow series:

$$T(s) = \sum_{n=0}^{\infty} Q_n s^{j+4}, . \quad (9)$$

**Step 3:** Utilizing the Elzaki inverse transformation for Equation (9), we derive the transformed expression, as follows:

$$E^{-1}\left[\sum_{i=0}^{\infty} Q_n s^{j+4}\right] = E^{-1}\left[\sum_{i=0}^{\infty} Y_n(n^2+3n+2)s^{j+4}\right] = \sum_{n=0}^{\infty}\frac{Y_n}{n!}t^{n+2} = Yt^2e^t, \quad (10)$$

**Step 4:** The Decrypted audio $X = Y -$ Lorenz Chaotic Key.

## 5. Results and Simulation Analysis

An audio encryption method should be capable of converting any audio into a cipher audio and then returning to the original plain audio during the decryption process. presents the results of the proposed methodology in Fig. 7. We will apply the Lorenz Chaotic key to the original audios and we will get the pre-encrypted audios for Piano, Car_Horn, and Human, as shown in Figures 7(d), 7(e) and 7(f). After that, we applied the Elzaki transformation to the pre-encrypted audios to get the encrypted audios as shown in Figures 7(g), 7(h) and 7(i). From Figure (7), it is clear that no one can guess anything about the original signal from the encrypted signal. The audio signal can be recovered exactly (see Fig. 7(j), 7(k) and 7(l)) by using the actual key. However, it cannot find the original signal using a slightly modified key. Therefore, the proposed method can properly encrypt and decrypt the audio signals. To analyze the performance of the proposed method, different performance metrics like entropy, mean square error (MSE), Peak Signal-to-Noise Ratio (PSNR), correlation coefficient, histogram analysis, and computational time analysis are used.

We will implement the intended approach using MATLAB R2021a. The implementation occurred on a personal computer system, equipped with an Intel(R) Core(TM) i5-1135G7 CPU running at a clock speed of 2.42 GHz. The computational unit is equipped with a spacious 1 terabyte hard disc drive for storage and a significant 8 GB of RAM to enable effective memory operations. We conducted this implementation using the Windows 11 Pro operating system.

**5.1 Statistical analysis.**

To verify that the proposed encryption cryptosystem was reliable, we undertook a number of statistical calculations. This report presents comprehensive data about the robustness and efficiency of the encryption scheme as shown by our tests in the field. Moreover, statistical research delves into the study of encryption outputs, analyzing audio signal value distributions, correlation coefficients, and error metrics including mean squared error (MSE) and peak signal-to-noise ratio (PSNR). These tests provide important information on how to ensure audio security and quality during the encryption and decryption techniques, as well as significant insights into the dependability and security of the encrypted data.

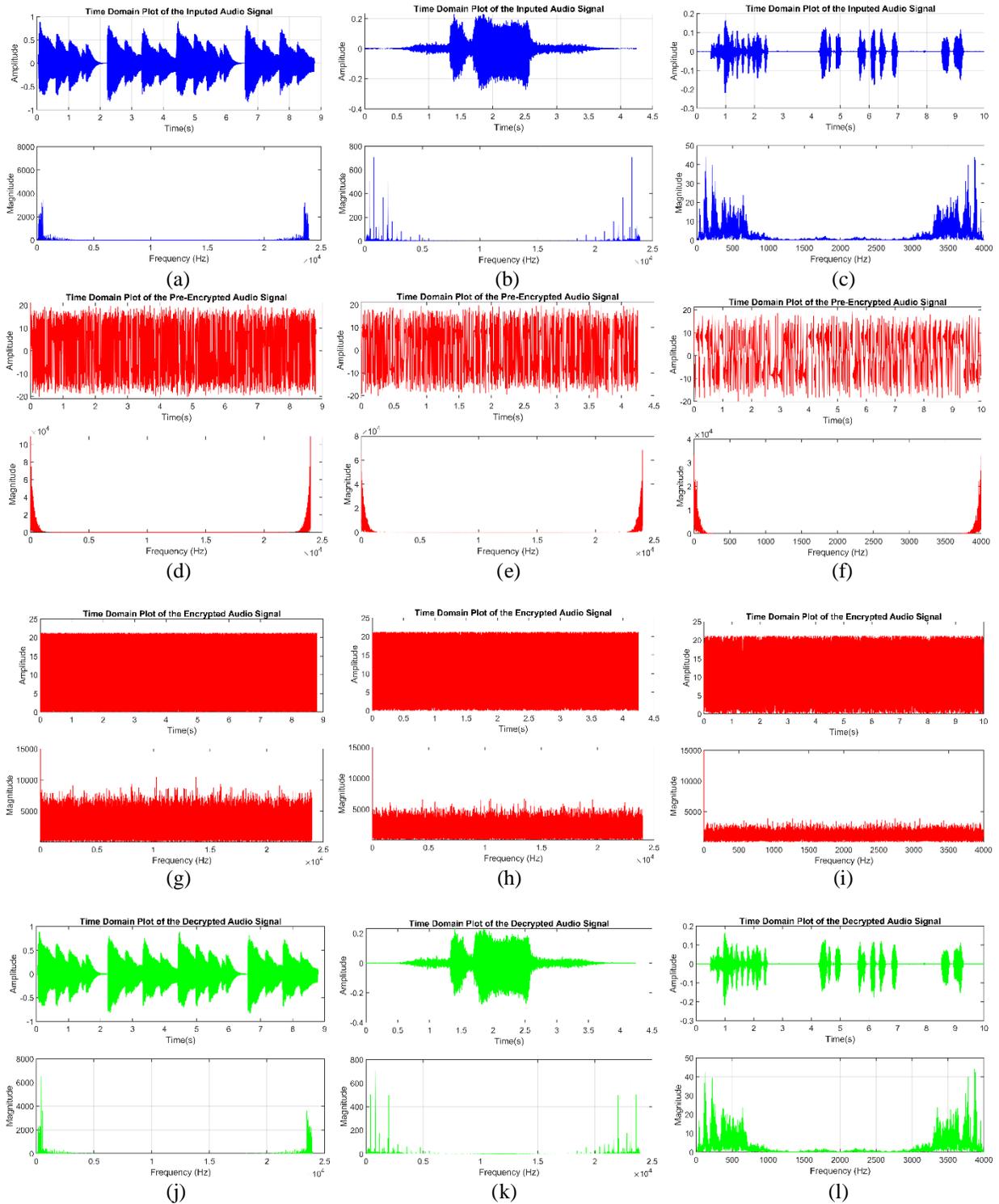

**Figure 7:** Results of audio pre-encryption, encryption and decryption for (a) Piano.wav plain audio; (b) Car_Horn.wav plain audio; (c) Human.wav plain audio; (d) Pre-encrypted Piano.wav; (e) Pre-encrypted Car_Horn.wav; (f) Pre-encrypted Human.wav; (g) Piano.wav encrypted audio; (h) Car_Horn.wav encrypted audio; (i) Encrypted Human.wav; (j) Piano.wav decrypted audio; (k) Car_Horn.wav decrypted audio; (l) Decrypted audio for Human.wav.

We just used the $x$-component of the Lorenz Chaotic Key for (d), (e), and (f), and we also used the Elzaki transformation for (g), (h), and (i) after applying the Lorenz Chaotic Key.

## 5.1.1 Histogram analysis

The histogram depicted in Figure 8 provides a visual representation of audio signal distributions across various stages: pre-encryption, encryption, and decryption. It showcases the amplitude levels for Piano.wav (a), Car_Horn.wav (b), and Human.wav (c) in their original state, alongside their pre-encrypted forms using the x-component of the Lorenz Chaotic Key (d, e, f). Furthermore, it displays the encrypted versions of these audio files (g, h, i) after applying both the Lorenz Chaotic Key and the Elzaki transformation.

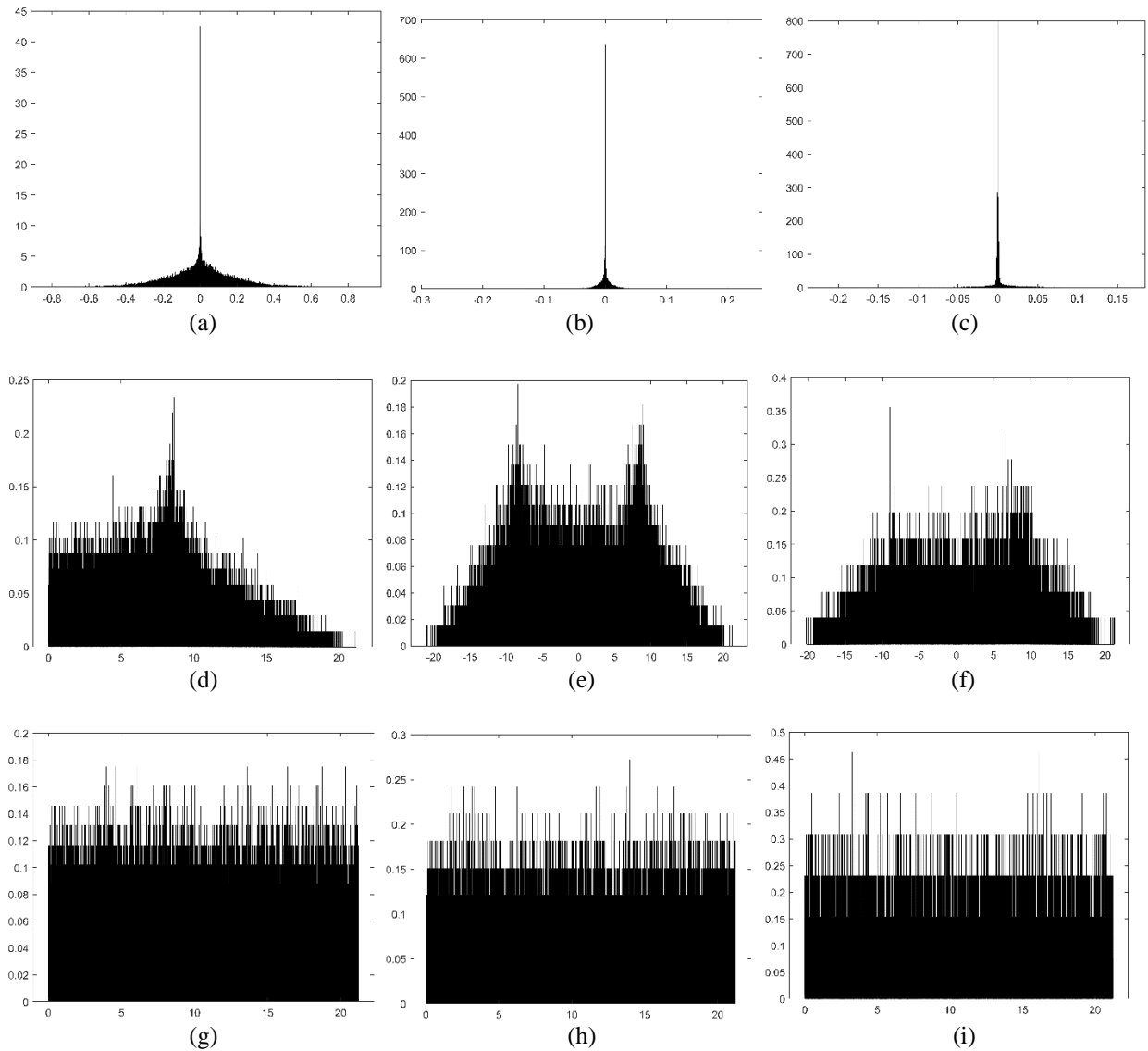

**Figure 8:** This histogram illustrates the distribution of audio signals at different stages: pre-encryption, encryption, and decryption. It includes the plain audio for Piano.wav (a), Car_Horn.wav (b), and Human.wav (c). Additionally, it displays the pre-encrypted audio for Piano.wav (d), Car_Horn.wav (e), and Human.wav (f) using the x-component of the Lorenz Chaotic Key. Moreover, it shows the encrypted audio for Piano.wav (g), Car_Horn.wav (h), and Human.wav (i) after initially applying the Lorenz Chaotic Key and then employing the Elzaki transformation.

This histogram analysis is critical for understanding how audio signals evolve through the encryption process, particularly concerning the prevention of information leakage. By examining the histograms of plain audio in Figure (8), it becomes evident that the original amplitude levels are preserved. However, the histograms for encrypted audio, also depicted in Figure 8, demonstrate a notable absence of identifiable amplitude patterns, indicating successful encryption.

Ensuring a uniform histogram distribution is paramount for fair energy allocation. Figure (8) shows how the encrypted audio histograms exhibit a smoother and more consistent pattern compared to their plain counterparts. This observation highlights the effectiveness of the proposed encryption approach in bolstering cryptographic security, as evidenced by the enhanced confidentiality of the audio signals.

### 5.1.2 Correlation Coefficient Analysis

A suitable encryption method should stop the correlation between nearby audio samples. Spreading sample values evenly across the intensity range and ensuring almost no correlation in the encrypted audio can achieve this. Figure 3, with a sample audio clip, illustrates this. The first column shows sample distributions in the plain audio, the second column shows the encrypted audio's correlation patterns, and the third column shows the decrypted audio. The results highlight successful decorrelation in the encrypted audio. Table 1 provides correlation coefficients for some audio samples.

Here's how to compute the correlation coefficient:

$$\rho(X,Y) = \frac{cov(X,Y)}{\sqrt{D(X)}\sqrt{D(Y)}}, \tag{11}$$

$$\bar{X} = \frac{1}{N}\sum_{i=1}^{N} X_i, \bar{Y} = \frac{1}{N}\sum_{i=1}^{N} Y_i,$$

$$D(X) = \frac{1}{N}\sum_{i=1}^{N}(X_i - \bar{X})^2, D(Y) = \frac{1}{N}\sum_{i=1}^{N}(Y_i - \bar{Y})^2$$

$$cov(X,Y) = \frac{1}{N}\sum_{i=1}^{N}\left((X_i - \bar{X})(Y_i - \bar{Y})\right),$$

The correlation coefficient values, represented as $\rho(x,y)$, range from $-1$ to $1$. When $\rho(x,y)$ is close to 1, it indicates a strong connection between neighboring pixels. A value of $\rho(X,Y)$ near 0 signifies no correlation, while negative values suggest an inverse or flipped relationship between the series. Correlation analysis of horizontal shift for Piano, Car_Horn, and Human audio signals at different stages of the encryption process is shown in Figure (9). In Table 1, we display the correlation coefficients for both unencrypted and encrypted audio data. To demonstrate the effectiveness of our proposed method, we performed a comparative assessment using Table 1, with the Piano, Car_Horn, and Human audios as our reference [6].

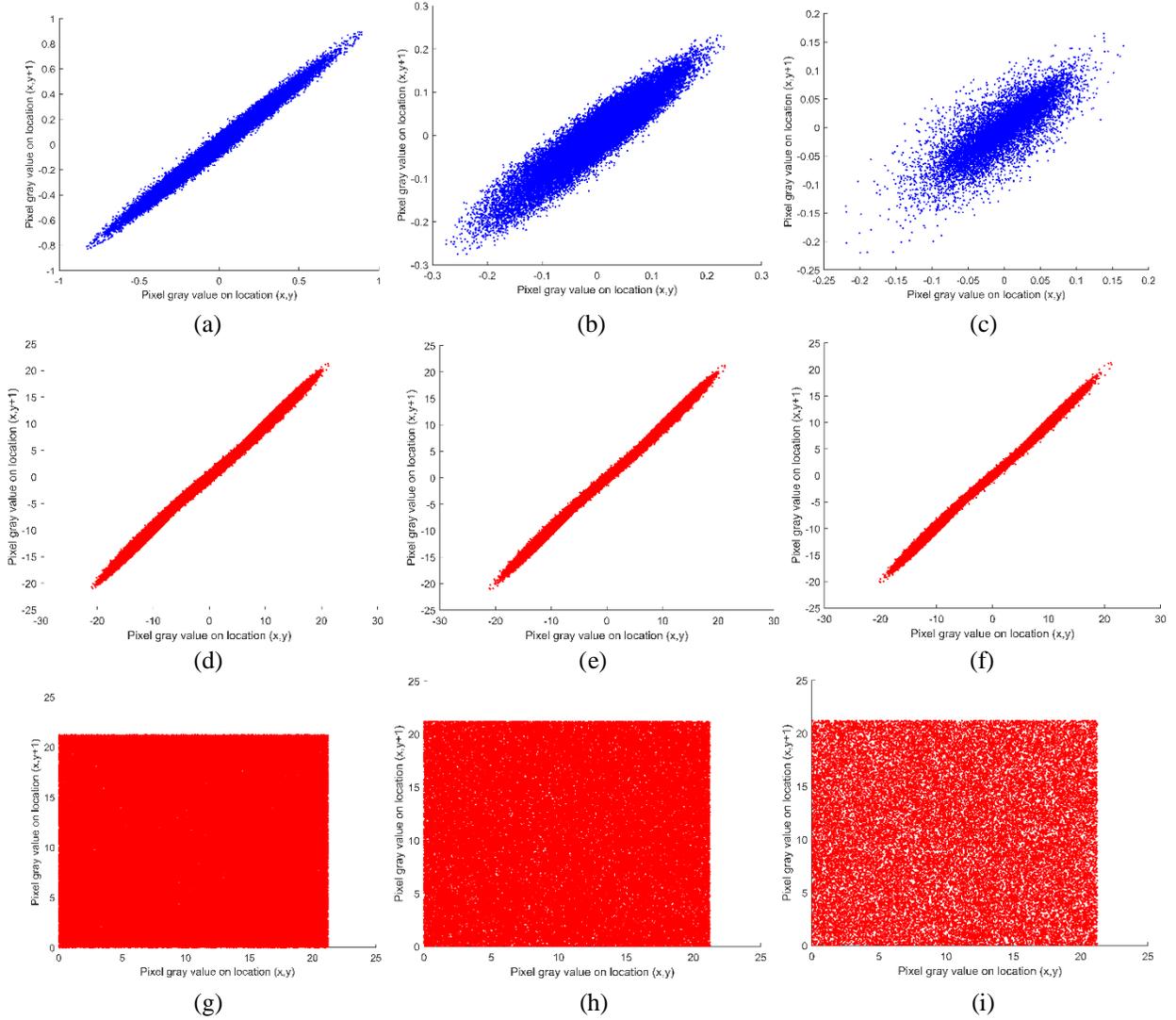

**Figure 9:** Correlation analysis of horizontal shift for various audio signals at different stages of the encryption process is shown. It includes the original audio for Piano.wav (a), Car_Horn.wav (b), and Human.wav (c). Additionally, it displays the pre-encrypted audio for Piano.wav (d), Car_Horn.wav (e), and Human.wav (f) using the x-component of the Lorenz Chaotic Key. Finally, it presents the fully encrypted audio for Piano.wav (g), Car_Horn.wav (h), and Human.wav (i) after applying the Lorenz Chaotic Key and then the Elzaki transformation.

**TABLE 1:** Correlation coefficients of horizontal shift for audio signals.

| Audios | Components of Lorenz Chaotic Key | Correlation Coefficients of Horizontal Shift | | |
|---|---|---|---|---|
| | | Original Audio | Pre-Encrypted Audio | Encrypted Audio |
| Piano | $x$ | 0.9925 | 0.9987 | −0.0014 |
| | $y$ | 0.9925 | 0.9976 | −0.0003 |
| | $z$ | 0.9925 | 0.9955 | 0.0023 |
| Car_Horn | $x$ | 0.9334 | 0.9987 | 0.0068 |
| | $y$ | 0.9334 | 0.9976 | 0.0057 |
| | $z$ | 0.9334 | 0.9955 | 0.0004 |
| Human | $x$ | 0.8025 | 0.9986 | −0.0014 |
| | $y$ | 0.8025 | 0.9975 | −0.0022 |
| | $z$ | 0.8025 | 0.9954 | −0.0025 |
| Ref. [11] | | 0.9936 | | -0.1578 |
| Ref. [12] | | 0.9970 | | 0.0133 |

Table 1 displays correlation coefficients that illustrate the correlation between horizontally shifted versions of various audio signals (piano, carhorn, and human) across three stages: original audio, pre-encrypted audio, and encrypted audio. The original audio correlation coefficients are consistently high for the audio signals; this indicates a strong relationship between the original audio signals. The pre-encrypted audio also has high correlation coefficients across all audio types and Lorenz components, with values between 0.9954 and 0.9987. This suggests that the structure of the original audio signals is kept during the pre-encryption process. However, the fully encrypted audio correlation coefficients are close to zero, ranging from -0.0025 to 0.0068, showing that encryption effectively removes the correlation between the audio signals and the Lorenz chaotic key components. This disruption means that the encrypted audio appears as noise to the chaotic system, indicating successful encryption that obscures the original signal structure.

**TABLE 2:** Correlation coefficients for certain audios.

| Audios | Components of Lorenz Chaotic Key | Correlation coefficients between original and encrypted audio. | Correlation coefficients between original and decrypted audio. |
|---|---|---|---|
| Piano | $x$ | 0.0005 | 1 |
|  | $y$ | -0.0010 | 1 |
|  | $z$ | -0.0015 | 1 |
| Car_Horn | $x$ | -0.0032 | 1 |
|  | $y$ | -0.0019 | 1 |
|  | $z$ | 0.0056 | 1 |
| Human | $x$ | -0.0026 | 1 |
|  | $y$ | 0.0038 | 1 |
|  | $z$ | 0.0069 | 1 |
| Ref. [13] |  | 0.0263 |  |
| Ref. [14] |  | 0.0263 |  |
| Ref. [15] |  | 0.0119 |  |

Table 2 shows the correlation coefficients between the original and encrypted audio and the original and decrypted audio for different audios with respect to the components of the Lorenz Chaotic Key. The correlation coefficients between the original and encrypted audio are near zero, indicating effective encryption that disrupts the original signal structure. In contrast, the correlation coefficients between the original and decrypted audio are all 1, demonstrating perfect recovery of the original audio after decryption. This highlights the effectiveness of the encryption and decryption processes in maintaining audio integrity while ensuring security. References[21,22] show a correlation coefficient close to zero between the original audio and the encrypted audio. This is much higher than the near-zero values we saw, which means the encryption is not as good. Reference[23] has a correlation coefficient of 0.0119, which is lower than those in References[21,22], but still higher than the almost-zero values of our proposed method, which can be seen in Table 2. This means that our proposed method is better at messing up the original audio signal.

## 5.2 Spectrogram Plotting

The spectrogram analysis provides a crucial method for examining audio signals, focusing on their frequency distribution over time. By comparing plain and encrypted files, we gain insight into the impact of encryption and can evaluate the efficacy of the proposed audio encryption algorithm.

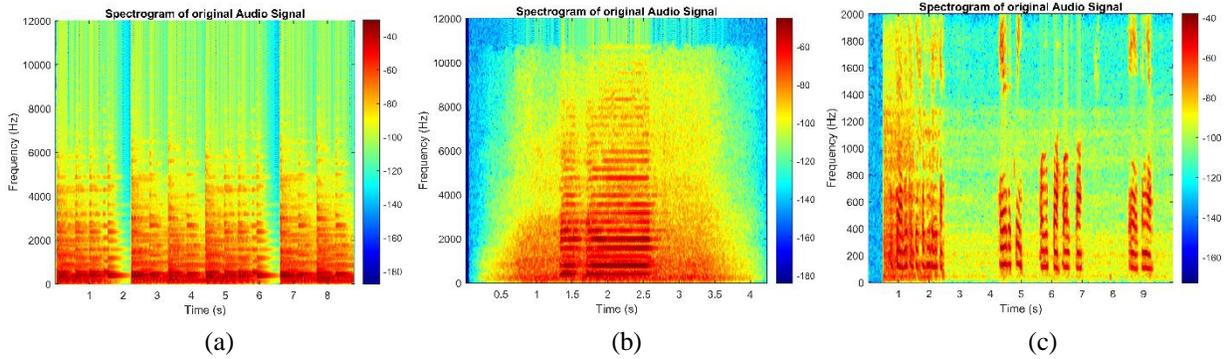

(a)           (b)           (c)

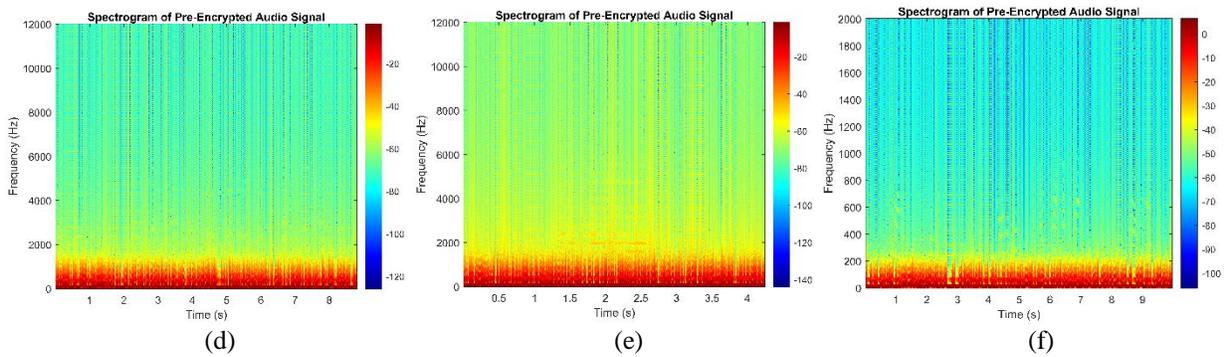

(d)           (e)           (f)

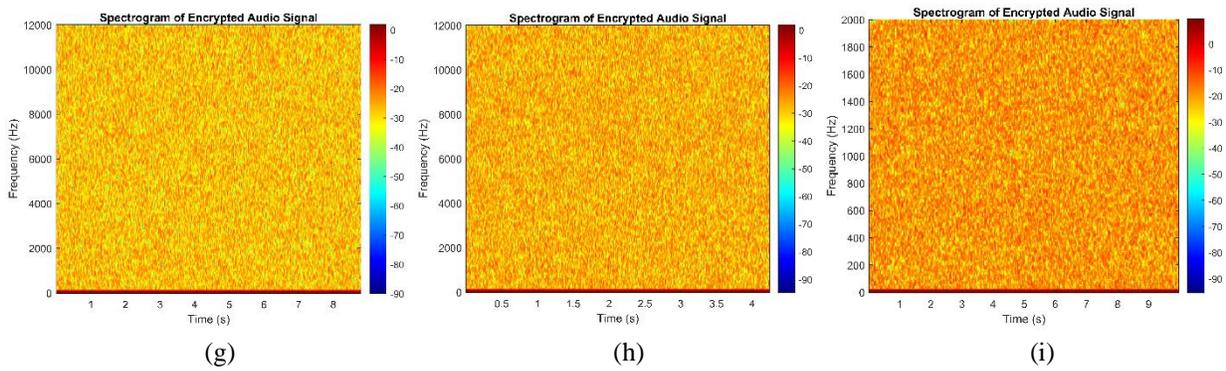

(g)           (h)           (i)

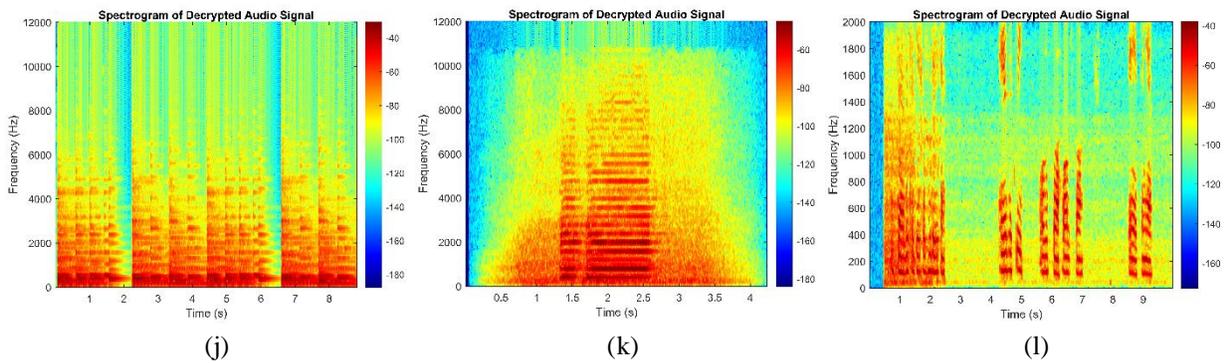

(j)           (k)           (l)

**Figure 10:** Spectrogram graphs of audio signals at different stages: pre-encryption, encryption, and decryption. It includes the original audio for Piano.wav (a), Car_Horn.wav (b), and Human.wav (c). It also shows the pre-encrypted audio for Piano.wav (d), Car_Horn.wav (e), and Human.wav (f) using the x-component of the Lorenz Chaotic Key. Additionally, it displays the encrypted audio for Piano.wav (g), Car_Horn.wav (h), and Human.wav (i) after applying the Lorenz Chaotic Key followed by the Elzaki transformation. Finally, it presents the decrypted audio for Piano.wav (j), Car_Horn.wav (k), and Human.wav (l).

In Figure 10, we observe spectrogram graphs illustrating audio signals at various stages: pre-encryption, encryption, and decryption. We depict the original audio files for Piano.wav (a), Car_Horn.wav (b), and Human.wav (c), followed by their respective pre-encrypted versions using the x-component of the Lorenz Chaotic Key (d, e, f). Subsequently, the encrypted audio files (g, h, i) are displayed, showing the effects of both the Lorenz Chaotic Key and the Elzaki transformation. Finally, the decrypted audio files (j, k, l) are presented. The spectrogram plots offer valuable insights into the encryption process. In particular, they demonstrate the obscuration of the original signals' frequency components in the encrypted files, signifying the loss of the original signal's frequency content. This result shows that the proposed algorithm is very good at encryption, as shown by the big difference in the spectrogram between plain and encrypted audio files.

### 5.3 Entropy Analysis

The rule for computing entropy, often referred to as Shannon's entropy, is a fundamental concept in information theory. It quantifies a random variable's uncertainty, or randomness. Here's Shannon's entropy rule [3]:

$$H(X) = -\sum_{i=1}^{n} P(x_i) \log_2 P(x_i),$$

where $H(X)$ is the entropy of the random variable $X$, $P(x_i)$ is the probability of the occurrence of outcome $x_i$ and $n$ is the total number of possible outcomes. Shannon's entropy provides a powerful tool for quantifying uncertainty and randomness in various fields, including information theory, cryptography, and data analysis. It forms the basis for understanding and analyzing the properties of random variables and probability distributions.

**TABLE 4:** Information entropy analysis.

| Audios | Original Audio | Encrypted Audio | Decrypted Audio |
|---|---|---|---|
| Piano | 14.2966 | 17.6896 | 14.2966 |
| Car_Horn | 10.975 | 16.6375 | 10.975 |
| Human | 9.6897 | 15.2877 | 9.6897 |

Table 4 presents an information entropy analysis for original and encrypted audio using components $x$, $y$, and $z$ of the Lorenz Chaotic Key and applying the Elzaki transformation. Information entropy is used to measure the randomness or unpredictability of data. Table 4 shows that for piano, car horn, and human audio, the original entropy values increase significantly after encryption across all components. Therefore, the significant increase in entropy from the original sound to the encrypted sound indicates that the encryption process effectively adds randomness and unpredictability to the data, which is desirable for secure encryption. The similarity in entropy values between the encrypted and decrypted sounds suggests that the decryption process accurately recovers the original randomness introduced during encryption. This is crucial for ensuring the fidelity and security of the decryption process. Overall, the computed entropy values provide valuable insights into the level of randomness, uncertainty, and effectiveness of the encryption and decryption processes applied to the sound data.

### 5.4 Perceptual security

This section examines the visual comparison between the encrypted data and the original unencrypted audio. We perform two distinct evaluations to assess the extent to which the proposed encryption technology preserves visual quality. The tests include subjective and objective exams, which, when combined, provide a comprehensive review of the visual security of the encryption process[24].

An alternative method of assessment entails scrutinizing impartial criteria used to assess excellence, as illustrated in Table 3. The subsequent section provides a mathematical elucidation of these specific measures of inaccuracies. We encourage readers to explore the details in Table 4 to gain a thorough understanding of the results.

1) Mean Squared Error (MSE) is a crucial metric for evaluating the effectiveness of audio encryption algorithms. It measures the average squared difference between the original and encrypted audio signals, quantifying the distortion introduced during encryption.

$$\text{MSE} = \frac{\sum_{i=1}^{\text{length}(X)} (X_i - C_i)^2}{\text{length}(X)}, \qquad (12)$$

What actually are exactly the distinctions within the encrypted audio and the original audio? An increased mean squared error (MSE) indicates an additional perturbation to the original signal, indicating a higher level of encryption security. An encrypted audio stream with a smaller mean square error (MSE) may nevertheless be susceptible to security risks since it preserves more of the original signal's features. In the realm of audio encryption, a large mean signal error (MSE) is advantageous as it signifies that the encryption technique has successfully concealed the original audio, making it unrecognizable and bolstering the defense towards illegal entry.

2) Peak Signal-to-Noise Ratio (PSNR) is a critical measure used to evaluate the quality of audio encryption algorithms. PSNR quantifies the ratio between the maximum possible power of the original audio signal and the power of the noise introduced by encryption, expressed in decibels (dB). A higher PSNR value indicates that the encrypted audio retains more of the original signal's quality, meaning less distortion and noise.

$$\text{PSNR} = 10 \times \log_{10}\left(\frac{Max^2}{\text{MSE}}\right), \quad (13)$$

The greatest intensity of the recordings is indicated as Max. When it concerns audio encryption, a low Peak Signal-to-Noise Ratio (PSNR) is frequently chosen since it indicates that the original signal has been effectively hidden by the encryption, hence enhancing its protection from unauthorized decoding. However, it is crucial to achieve a balance, since having a PSNR that is too low will substantially damage the quality of the signal and impact the practicality of the decrypted audio. PSNR provides an important gauge for assessing the balance between security and sound quality in encryption methods.

**TABLE 3:** Quantitative metrics measuring the perceptual quality of the plain audio in comparison to the encrypted audio and the audio after decryption.

| Audios | Components of Lorenz Chaotic Key | Original plain audio with encrypted audio | | Original plain audio with decrypted audio. | |
|---|---|---|---|---|---|
| | | MSE | PSNR (dB) | MSE | PSNR |
| Piano | $x$ | 149.7306 | 4.78 | 0 | ∞ |
| | $y$ | 278.9325 | 4.78 | 0 | ∞ |
| | $z$ | 954.4959 | 4.78 | 0 | ∞ |
| Car_Horn | $x$ | 150.6632 | 4.75 | 0 | ∞ |
| | $y$ | 280.1603 | 4.76 | 0 | ∞ |
| | $z$ | 956.3056 | 4.77 | 0 | ∞ |
| Human | $x$ | 149.8758 | 4.78 | 0 | ∞ |
| | $y$ | 280.1483 | 4.76 | 0 | ∞ |
| | $z$ | 949.5719 | 4.80 | 0 | ∞ |

Table 3 compares the quality of the original plain audio to encrypted audio and audio that has been decrypted. It does this by using quantitative metrics, such as mean squared error (MSE) and peak signal-to-noise ratio (PSNR), both measured in decibels (dB). We perform the analysis on three types of audio samples: piano, carhorn, and human, using components x, y, and z of the Lorenz Chaotic Key. For all types of audio, the MSE between the original and encrypted audio is notably high, indicating significant differences, while the corresponding PSNR values are low, around 4.75 to 4.80 dB, signifying perceptible noise. In contrast, the MSE between the original and decrypted audio is zero, and the PSNR is infinite, indicating a perfect reconstruction with no loss of quality after decryption.

1. **CONCLUSION**

This work provides an original technique for encryption audio employing the chaotic dynamics of the Lorenz system and the Elzaki transformation. The Lorenz chaotic system provides a cryptographic key that optimizes security by depending on its sensitive sensitivity on beginning conditions and unpredictable behavior. The Elzaki transformation, recognized for its remarkable transformative characteristics, enhances the security of audio data by changing the coefficients of the Maclaurin series. The encryption procedure successfully conceals the original audio signals, as shown by the statistical and correlation analysis. The encrypted audio exhibits a notably elevated level of entropy, suggesting enhanced levels of unpredictability and security. Moreover, the decrypted audio remains unchanged in terms of its original quality, devoid of any noticeable disruptions or loss of information, thereby affirming the reliability of the entire procedure.

Performance metrics such as Mean Squared Error (MSE) and Peak Signal-to-Noise Ratio (PSNR) demonstrate that the proposed technique successfully balances security and audio fidelity. The method's computational efficiency is proven by its implementation in MATLAB R2021a on a standard personal computer system. Conclusively, the suggested technique of encryption is a reliable and efficient means of safeguarding audio information, permitting exceptional security as well as conservation of audio integrity. Future research endeavors may examine the relevance of this method of study to different types of multimedia data and its incorporation into real-time communication platforms.

**AUTHOR DECLARATION**

I confirm that all the Figures and Tables in the manuscript are mine.

**ACKNOWLEDGMENTS**

The authors thank Koya University for their financial assistance.

.